\documentclass[11pt]{llncs}
\usepackage[a4paper,twoside=false,top=2.32in,bottom=2.69in,left=2.25in,right=2in]{geometry}
\usepackage{times}

\usepackage{amssymb}
\usepackage{graphicx}

\usepackage{xcolor}

\usepackage{url}

\urldef{\mailsa}\path|{carlos-emiliano.gonzalez-gallardo, eric.sanjuan, juan-manuel.torres}@univ-avignon.fr|
  
\newcommand{\keywords}[1]{\par\addvspace\baselineskip
\noindent\keywordname\enspace\ignorespaces#1}

\begin{document}

\mainmatter  

\title{Extending Text Informativeness Measures to Passage Interestingness Evaluation}
\subtitle{Language Model vs. Word Embedding}
\titlerunning{Extending Text Informativeness Measures to Passage Interestingness Evaluation}

%
%
\author{Carlos-Emiliano Gonz\'alez-Gallardo\inst{1}%
\and Eric SanJuan\inst{1}\and Juan-Manuel Torres-Moreno\inst{1,2}}
%


\institute{LIA, Universit\'e d'Avignon et des Pays de Vaucluse,\\
74 Rue Louis Pasteur, 84029 Avignon, France\and \'Ecole Polytechnique de Montr\'eal, 2900 Edouard Montpetit Blvd,\\
Montreal, QC H3T 1J4, Canada\\
\mailsa
}

%
%

\maketitle

\begin{abstract}
Standard informativeness measures used to evaluate Automatic Text Summarization mostly rely on n-gram overlapping between the automatic summary and the reference summaries.
These measures differ from the metric they use (cosine, ROUGE, Kullback-Leibler, Logarithm Similarity, etc.) and the bag of terms they consider (single words, word n-grams, entities, nuggets, etc.).
Recent word embedding approaches offer a continuous alternative to discrete approaches based on the presence/absence of a text unit. Informativeness measures have been extended to Focus Information Retrieval evaluation involving a user's information need represented by short queries.
In particular for the task of CLEF-INEX Tweet Contextualization, tweet contents have been considered as queries.
In this paper we define the concept of Interestingness as a generalization of Informativeness, whereby the information need is diverse and formalized as an unknown set of implicit queries.
We then study the ability of state of the art Informativeness measures to cope with this generalization.
Lately we show that with this new framework, standard word embeddings outperforms discrete measures only on uni-grams, however bi-grams seems to be a key point of interestingness evaluation.
Lastly we prove that the CLEF-INEX Tweet Contextualization 2012 Logarithm Similarity measure provides best results.
\end{abstract}

\section{Introduction}

Following Bellot \textit{et al.} in \cite{DBLP:journals/ipm/BellotMMST16}, we consider Informativeness in the context of Focused Retrieval (FR) \cite{DBLP:journals/sigir/KampsGT08}.
Given a query representing a user information need, a FR system returns a ranked list of short passages extracted from a document collection.
Then, a user reads a passage top to bottom and tag it as: 1) informative, 2) partially informative or 3) uninformative depending if all, only parts or no part of it contains useful information relevant to the query.
This information can be factual and explicitly linked to the query as in Question Answering (QA) or more abstract and provide some general background about the query. FR systems are evaluated according to the cumulative length of extracted informative passages (Precision) and their diversity (Recall).

Interestingness, by contrast,  is a much broader concept used in Data Mining as it is defined as  \textit{``the power of attracting or holding  attention''} and relates the ideas of lift and information gain used to mine very large sets of association rules between numerous features.
Mining interesting associations between features is a complex interactive process.
Unlike Information Retrieval (IR), in Interestingness  there is no precise query to initiate the search.

As expected, it has been shown that none of the numerous Interestingness measures proposed in Business Intelligence softwares allow to catch all associations that experts will consider as useful \cite{koh_interestingness_2008}.
Each measure allows to grasp only a facet on interestingness.
However, Hilderman \textit{et. al} defined in \cite{Hilderman:2003} five principles to be satisfied by these type of measures: minimal and maximal values, skewness, permutation invariance and transfer.

By reusing this concept of Interestingness, in the context of FR we define it as: \textit{a text passage that is clearly informative for some implicit user's information need.} More precisely; given a set of users and a set of passages that were considered interesting by at least one of these users, the task consists in finding new interesting passages not related to previous topics and implicit queries.
We test the ability of state of the art FR Informativeness metrics to deal with this specific new task. Therefore, in this paper we consider short factual passages, most of them are single sentences instead of complete summaries.
From a formal point of view we consider text units like words, word n-grams, terms or entities as attributes in \cite{Hilderman:2003}.
Passages $S$ to be ranked by decreasing interestingness  are then represented as sets of  unique tuples $(\omega, \omega(S))$  where  $\omega$ is an attribute and $\omega(S)$ its frequency.

The data we used consists of a large pool of $34,507$ passages extracted from the Wikipedia by state of the art FR systems \cite{DBLP:journals/ipm/BellotMMST16} over $67$ topics from Twitter.
Each passage real informativeness has been individually assessed by CLEF-INEX Tweet Contextualization\footnote{\url{http://inex.mmci.uni-saarland.de/tracks/qa/}} (TC@INEX) task organizers.
These assessments resulted in a reference score scaled between 0 and 2 depending on the relative length of the informative part and assessor's inter-agreement. 

Based on this extended dataset we analyze correlations between the reference Informativeness scores and a set of state of the art Informativeness automatic measures including F-score, Kull\-back-Leibler (KL) divergence, ROUGE, Logarithm Similarity (Log\-Sim) and cosine measures.
Because passages are extracted from the Wikipedia, we also consider the restriction of these measures to anchor texts inside passages.
Wikipedia anchors are references to related entities; anchors inside informative passages can be considered as informative nuggets annotated by Wikipedia contributors.

The rest of the paper is organized as follows. In \S \ref{sec:stateArt} we recall the state of the art about Informativeness measures, nugget based evaluation and Interestingness principles.
In \S \ref{sec:Method} we detail the method we used to evaluate the relative effectiveness of each metric when applied to passage informativeness and Interestingness evaluation.
\S \ref{sec:Results} presents and discuss the results. Finally, \S \ref{sec:Conclusion} concludes this paper.

\section{State of the Art}
\label{sec:stateArt}

Informativeness measures have been introduced in various contexts. Depending of the type of information unit that is used and the way this unit is represented, it is possible to separate them in language model approaches and continuous space approaches (word embeddings).

\subsection{Language Model Approaches}

Language  model (LM) approaches of Informativenesses are based on smoothed probabilities over all text units.
The most popular is ROUGE, which compares an automatically generated summary with a (human-produced) reference summary.
ROUGE-N is defined as the average number of common term n-grams in the summary and a set of $k\geq 5$ reference summaries \cite{Lin2004Rouge}.
In particular, ROUGE-1 computes the distribution of uni-grams; it counts the number of uni-grams (words or stems) that occur both in the reference summary and in the produced summary \cite{Torres2014}.

Apart from ROUGE-1, other variants of ROUGE are also available.
While ROUGE-1 (uni-grams) considers each term independently, ROUGE-2 (bi-grams) considers sequences of two terms.
Skip-grams, used in ROUGE-S, are pairs of consecutive terms with possible insertions between them; the number of skipped terms is a parameter of ROUGE-SU. 

ROUGE-2 and ROUGE-SU4 were used to evaluate the generated summaries in the Document Understanding Conference (DUC) in 2005 \cite{dang2005overview}.
ROUGE proved itself better correlating human judgments under readability assumption than classical cosine measures \cite{DBLP:conf/acl/RadevTSLBQCLD03}. 
Indeed, the various ROUGE variants were evaluated on the three years of DUC data in \cite{lin2004looking}, showing that some ROUGE versions are more appropriate for specific contexts. 

ROUGE implies that reference summaries have been built by humans.
In the case of a very large number of documents, it is easier to apply a measure that can be used automatically to compare the content of the produced summary with the one of the full set of documents.
In this framework, measures such as KL divergence and Jensen-Shannon (JS) probability distribution divergence used in the Text Analysis Conference (TAC) \cite{dang2007overview} compare the distributions of words or word sequences between the produced summary and the original documents.
In this approach, Informativeness relies on complex hided word distributions and not on specific units.

In addition, it is shown that JS probability distribution divergence is correlated in the same way than ROUGE when used to evaluate summaries built by passage extraction from the documents to be summarized \cite{LouisN:09,DBLP:conf/coling/SaggionMCSV10}.
JS is the symmetric variant of the KL divergence used in the LM for IR and Latent Dirichlet Allocation to reveal latent concepts.
Both measures aim at calculating the similarity between two smoothed conditional probabilistic distributions of terms  $P(w|R)$ and $P(w|S)$ where $R$ is a textual reference and $S$ a short summary. 

We can also cite the LogSim measure used in TC@INEX 2012 task \cite{DBLP:conf/clef/SanJuanMTBM12}. This measure was introduced because 1) there were no reference summaries produced by humans, 2) the automatically produced summaries were of various sizes and 3) some automatically produced summaries  were too short. 
Although these various measures have been introduced to evaluate similar tasks as Automatic Text Summarization (ATS), they are applied in different contexts. 
Moreover, all of them can be sensitive to the text unit that is being used.
The influence of the type of text unit (uni-gram, bi-gram, skip-gram, etc.) has been evaluated for ROUGE in \cite{lin2004looking} as they correspond to the ROUGE variants.

Informativeness measures can be classified into three main families:
The most common approach of Informativeness likelihood in IR relies on probabilistic KL divergence.
In automatic summarization evaluation the most common approaches are $F_\beta$ and ROUGE similarity scores based on n-grams.
This two families rely on exact term overlapping.
A third family created by word embedding representations has recently introduced a vectorial approach less dependent on explicit term overlapping.

More formally, let $\Omega$ be a type of text unit and $S$ a sentence which Informativeness is to be evaluated against a textual reference $R$, we use the following three discrete metrics and one continuous approach.

\subsection*{Kullback-Leibler (KL) Divergence}

We implement it as the expectation based on the reference $R$ of the $log$ difference between normalized frequencies in $R$ and smoothed probabilities over $S$.
We fix the smoothing parameter $\mu$  at its minimal value ($\mu=1$).
This divergence is not normalized. 

\begin{equation}
D_{KL}(R||S)= \sum_{\omega\in\Omega(R)} 
\ln\left(
\frac{P(\omega|R).(|S|+1)}{P(\omega|S).|S|+P(\omega|\Omega)}\right).P(\omega|R)
\end{equation}

\subsection*{Logarithm Similarity (LogSim)}

Like the KL divergence, this is also an expectation based on the reference $R$ but of a normalized similarity that is only defined over $\Omega(S)\cap\Omega(R)$ \cite{DBLP:conf/clef/SanJuanMTBM12}. 

\begin{equation}
LS(S||R)=\sum_{\omega\in\Omega(S)\cap\Omega(R)} 
e^{-\left|\ln\left(
\frac{L_R(\omega,S)}
{L_R(\omega,R)}
\right)\right|}. P(\omega|R)\\
\end{equation}
where,
\begin{equation}
L_R(\omega,X)=\ln(1+P(\omega|X).|R|)
\end{equation}

\subsection*{$F_\beta$ and ROUGE Scores}

$F_\beta$ measures the harmonic mean between Precision ($p$) and Recall ($r$). 
It is normally expressed as

\begin{equation}
F_\beta=\frac{(\beta^2+1) \times p \times r}{\beta^2 \times p + r}
\label{formula:f_b}
\end{equation}

\noindent where $\beta$ is the factor that controls the relative emphasis between $p$ and $r$. 
If $\beta = 1$, then it is possible to rewrite Equation \ref{formula:f_b} as: 

\begin{equation}
	F_1=2\times\frac{p \times r}{p + r}
\label{formula:f_1}
\end{equation}

F1-score ($F_1$) is the most common normalized set theoretic similarity giving equal emphasis to $p$ and $r$. 
To represent $F_1$ in terms of $\Omega$, $R$ and $S$, lets first rewrite $p$ and $r$ as: 

\begin{equation}
p=\frac{|\Omega(S) \cap \Omega(R)|}{|\Omega(S)|}
\end{equation}
\begin{equation}
r=\frac{|\Omega(S) \cap \Omega(R)|}{|\Omega(R)|}
\end{equation}
\noindent finally Equation \ref{formula:f_1} is redefined as:
\begin{equation}
F_1(S|R)=2\times\frac{|\Omega(S)\cap\Omega(R)|}{|\Omega(S)|+|\Omega(R)|}
\end{equation}

As explained in \cite{Lin2004Rouge}, the idea behind all ROUGE metrics  is to automatically determine the quality of a candidate summary comparing it with reference summaries written by humans. The quality is obtained by comparing the number of overlapping n-grams, such as word sequences or word pairs, between the candidate and a set of reference summaries. 

ROUGE-N is defined as the average number of common n-grams between the candidate summary and a set of reference summaries:

\begin{equation}
\textrm{ROUGE-N}= \frac{\sum_{\omega\in\Omega(R)} Count_{match}(\omega)}{\sum_{\omega\in\Omega(R)} Count(\omega)}
\end{equation}

\noindent where $Count(\omega)$ is the frequency of the $\omega's$ n-gram and $Count_{match}$ is the co-occurring frequency of the $\omega's$ n-gram in $R$ and in the candidate summary $S$.
All ROUGE variants base their functionality in the lexical similarities between the candidate and the reference summaries, this is a problem when the candidate summary do not share the same words of the references; that is the case of abstractive summaries and summaries with a significant amount of paraphrasing. 

\subsection*{Continuous Space Approach}

Classic approaches of Natural Language Processing transform the words of a dataset into a bag of words representation that leads to sparse vectors and semantic information loose.
In a dataset containing $n$ different words, each word will be represented in a one-hot vector that is absolutely independent of the rest of the words in the dataset. 

Table \ref{table:one-hot} shows the one-hot vectors of a fictitious dataset with 10,000 different words. It is true that some kind of relation exists between $tiger \leftrightarrow cat$ and $wolf \leftrightarrow dog$ but with the one-hot vectors this relationship is impossible to  maintain. The restriction can be overcome with word embeddings.

\begin{table}
  \centering
  \caption{One-hot encoding vectors example}  
  \begin{tabular}{lcc}
  \hline
  ID &word & One-hot \\
  \hline
  1 & tiger & $[1\ 0\ 0\ 0\ 0\ ... 0\ 0]$ \\
    2 &cat &   $[0\ 1\ 0\ 0\ 0\ ... 0\ 0]$ \\
    ...& ...& ...\\
  9999 & wolf &  $[0\ 0\ 0\ 0\ 0\ ... 1\ 0]$ \\
  10000 & dog &   $[0\ 0\ 0\ 0\ 0\ ... 0\ 1]$ \\
  \hline
\end{tabular}
\label{table:one-hot}
\end{table}

Word embeddings are another way to represent words within a dataset.
In this representation, vectors are capable to maintain the relationship between words in the dataset following the distributional semantics hypothesis \cite{harris1954distributional}: \textit{words that appear in the same contexts share similar meanings}.

Word2vec is a popular Neural Network embedding model described in \cite{mikolov2013linguistic}.
It aims to map the vocabulary of a dataset into a multidimensional vector space in which the distance between the projections corresponds to the semantic similarity between them \cite{ng2015better}. 


Two different model approaches are proposed in word2vec: Continuous Bag-of-Words (CBOW) and Continuous Skip-gram.
The first one predicts a target word based on the context, while the second one predicts a context given a target word \cite{Mikolov2013_Efficient}.
In both cases the size of the embeddings is defined by the size of the projection layer. 


Learning the output vectors of the neural network is a very expensive task; so in order to increase the computation efficiency of the training process, it is necessary to limit the number of output vectors that are updated for each training instance. In \cite{rong2014Word2vec} two different methods are proposed: Hierarchical Softmax and Negative Sampling.

\begin{itemize}
\item \textbf{Hierarchical Softmax \cite{morin2005hierarchical}:} It represents the output layer of the neural network in a binary tree shape where each leaf of the tree corresponds to each one of the $V$ words of the vocabulary within the dataset and each node represents the relative probabilities of its child nodes.
Given the nature of the binary tree shape, there exists just one path from the root of the tree to each one of the leafs; making it possible to use this path to estimate the probability of each one of the words (leafs) \cite{mikolov2013distributed,rong2014Word2vec}.

\item \textbf{Negative Sampling \cite{mikolov2013distributed}:} The goal is to reduce the amount of output vectors that have to be updated during each iteration of the training process.
Instead of using all the samples during the loss function evaluation, just a small sample of them is taken into account.
\end{itemize}

\noindent As proposed in \cite{mikolov2013distributed}, it is possible to  combine words by an element-wise addition of their embeddings.
Given a document D of length $n$  represented by the set of word embeddings $D = \{\overline{v}_1, \overline{v}_2, ..., \overline{v}_n\} $ where $|\overline{v}_i| = m$; a simple way to represent D with a unique embedding in terms of $D$ is to add each component $j$ of each embedding $\overline{v}_i$ in $D$ to obtain a unique vector  $\overline{d}$ of length $m$.

To measure the similarity between the vector of a reference document ($\overline{d}_R$) and the one of a proposed sentence ($\overline{d}_S$) we calculate the cosine similarity between the two vectors as:
\begin{equation}
cos_{RS}(\theta) = \frac{\overline{d}_R\cdot\overline{d}_S}{||\overline{d}_R||\cdot||\overline{d}_S||}
\end{equation}

\subsection{Nugget Based Evaluation}

LM approaches of Informativeness are based on smoothed probabilities over all text units, but it is possible to use only a subset of them (nuggets).

In the context of QA, Dang \textit{et al.} defined a nugget as an informative text unit (ITU) about the target that is interesting; atomicity being linked to the fact a binary decision can be made on the relevance of the nugget to answer a question \cite{dang2007overview}.
This method makes it possible to consider documents that have not been evaluated to be labeled as relevant or not relevant (simply because they contain relevant nuggets or not). 

It has been shown that real ITUs can be automatically extracted to convert textual references into a set of nuggets \cite{DBLP:conf/sigir/Ekstrand-AbuegPA13}.
This simplifies the complex problem of Informativeness evaluation providing a method  to measure the proximity between two sets of ITUs.
For that, standard Precision-Recall or Pyramid measures can be used if nuggets are unambiguous entities \cite{DBLP:conf/sigir/LinZ07} and more sophisticated nugget score metrics based on shingles if not \cite{DBLP:conf/wsdm/PavluRGA12}.

We refer the reader to these publications \cite{dang2007overview,DBLP:conf/sigir/Ekstrand-AbuegPA13,DBLP:conf/sigir/LinZ07,DBLP:conf/wsdm/PavluRGA12} for advanced and non trivial technical details.
Here we shall point out that the original Pyramid method relies on human evaluation to: 

\begin{enumerate}
\item Identify in text summaries all short sequences of words that are relevant to some query or question.
\item Supervise the clustering of these units into coherent features that allow to compute some informativeness score.  
\end{enumerate}

The global idea is that Informativeness relies on the presence or absence of some specific text units and can be then evaluated based on their counting. 
Pavlu \textit{et al.} share this same idea even though they try to automatize the process of selecting and identifying these ITUs \cite{DBLP:conf/wsdm/PavluRGA12}. 

\subsection{Interestingness Principles}
For the purpose of this work, we consider sentences as sets of items (nuggets, words or word n-grams).  There is a wide range of Interestingness measures but they all combine the three following properties \cite{Hilderman:2003}:

\begin{enumerate}
\item Diversity that can rely on concentration (the sentence reveals an important  information) or on dispersion (the sentence links several different entities).  
\item Permutation invariance: the order of the items has no impact on the overall score.
\item Transfer: a sentence with few highly informative items should be considered more interesting than a long sentence with less informative items. 
\end{enumerate}

Given a text reference, only the LogSim measure covers all three properties.
KL measure, due to smoothing, does not fulfill (1); while $F_\beta$ measure does not fulfill (3) since it favors long sentences with large overlaps.

\section{Method}
\label{sec:Method}

\subsection{From Informativeness to Interestingness}

We formally define short passage Informativeness evaluation as a ternary relation between a set of topics $T$, a subset of short text extracts $P$ from a large document resource and a set $S$ of graded scores such that top ranked passages contain certainly relevant information about the related topic or its background.

By contrast, we define short passage Interestingness evaluation as a projection of interestingness over content and scores. Therefore a binary relation exist between a set of short text passages and a graded score such that top ranked passages are informative for some unknown topic. 

\subsection{Dataset}

We consider the data collection from the TC@INEX 2012 task \cite{DBLP:conf/clef/SanJuanMTBM12} and state of the art measures to automatically evaluate summary Informativeness and Interestingness.
The participants to this task had to provide a summary composed of extracted passages from Wikipedia that should contextualize a tweet by revealing its implicit background and providing some factual explanations about related concepts. 
During 2012 and 2013, tweets were collected from non-personal Twitter accounts. 
In 2012 a total of 33 valid runs were submitted by 13 teams from 10 countries; while in 2013, 13 teams from 9 countries submitted 24 runs to the task \cite{bellot2013overview}.

\subsection*{Textual Relevance Judgments}

Reusable assessments to evaluate text informativeness have been one of the main goals of TC@INEX tracks; all the collections that have been built are indeed freely available\footnote{\url{http://tc.talne.eu}}.
This has been performed by organizers on a pool of 63 topics (tweets) in 2012, and 70 topics in 2013.

Assessments consist in textual relevance judgments (t-rels) to be used for content comparison with automatically built summaries.
Since summaries returned by participant systems are  composed of extracted passages from Wikipedia, the reference is built on a pool of these passages.

An important fact in the pre-processing of this pool is that passages starting and ending by the same 25 characters have been considered as duplicated; therefore passages in the reference are unique, but short sub-passages could appear twice in longer ones.
Moreover, since for each topic all passages from all participants have been merged and displayed to the assessor in alphabetical order, each passage informativeness has been evaluated independently from others, even in the same summary.
This results in a reference highly  redundant at the level of noun phrases and consequently all types of ITUs that can be extracted.

The soundness of the pooling procedure was verified during the TC@INEX 2011 campaign over a corpus extracted from the New York Times\footnote{\url{https://www.nytimes.com}}  \cite{DBLP:conf/inex/2011,DBLP:conf/inex/SanJuanMTBM11}.
Indeed, topics in 2011 were only tweets coming from the New York Times, and a full article was associated to each tweet.
To check that the resulting pool of relevant answers was sound, a second automatic evaluation for informativeness of summaries had then been carried out, taking as the reference the textual content of the article.
None of the participants had reported having used this information available on the New York Times website.
Both rankings, one based on submitted run pooling and the second one based on New York Times articles, appeared to be highly correlated. Pearson’s product-moment correlation $= 88\%$ and $p$-value$< 10^{-6}$.

The TC@INEX 2012 task collection provides:
\begin{itemize}
\item A set of $63$ different topics: A topic is a short sentence (a tweet) that is used by participants to build a query-driven summary. Each summary should be 500 words-long or less and be supposed to be built by sentence extraction from Wikipedia;
\item A set of $36$ runs: a run consists of several summaries, one per topic which has to be built by one system from a participant;
\item For each summary produced by a participant, the passages that have been marked as informative (and thus the ones that are considered as non informative) by human assessors. We have a set of $34,507$ assessed passages, either informative or non-informative.
\end{itemize}

From this test collection we extracted the set of passages from participants' automatic summaries that human evaluators have mar\-ked as informative regarding a topic.
Each topic was evaluated by two people including the one that chose the original tweet as the topic.
Therefore each passage obtained a graded score among $[0,1]\cup\{2\}$ as the relative total length of the passage that has been highlighted as informative by at least one evaluator.
In the case that the two evaluators agreed that the whole passage was informative, an score of $2$ was assigned.

Another resource was built using this data.
From each passage we extracted text units.
We considered in first place stems that were in turn used to build uni-grams, bi-grams and skip-grams; secondly, Wikipedia entities in anchor texts.
Stems were simply obtained using Porter's stemmer after a process of stop words removal; they correspond to uni-grams. Bi-grams were composed of two adjacent text units after stop words removal. 

If all passages that have been selected as informative are considered and separated by topic, it is possible to build a textual reference for each topic to apply state of the art Informativeness metrics. By contrast, by merging all the references we obtain a textual reference to evaluate Interestingness. 

We also focus on other type of token characteristics. We hypothesize that the measures  also can be sensitive to term distribution.
Indeed, it is likely that automatic summaries that contain highly frequent terms are less interesting for a user than a summary that contains less frequent and thus probably more informative terms.
In the same way, it is likely that summaries that contain name entities are more informative than summaries that do not contain any.
These phenomenons have not been studied in the literature justifying the purpose of this paper.
While in previous studies tokens are words or stems, in this paper we also study other types of tokens such as DBpedia entities.

To perform experiments with word embeddings we used four different word2vec models.
All of them with an embedding size of 300 dimensions and a negative sampling of 15 units. Table \ref{table:models_embeddings} shows the number of embeddings and the text unit used in each model.

\begin{table}
  \centering
  \caption{Word2vec models}  
  \begin{tabular}{rrc}
Model~~~~~~ & ~\# of embeddings~ & ~Text Unit\\
  \hline
  $\textrm{google}\_\textrm{news}_{1gram}$& 3,000,000~ & ~uni-grams \\
  $\textrm{clef}\_\textrm{inex}_{1gram}$&30,000~ & ~uni-grams \\
  $\textrm{clef}\_\textrm{inex}_{2gram}$&600,000~ & ~bi-grams \\
  $\textrm{wiki}_{2gram}$&10,000,000~ & ~bi-grams
  \end{tabular}
  \label{table:models_embeddings}
\end{table}

The $\textrm{google}\_\textrm{news}_{1gram}$ mo\-del was the same used by \cite{ng2015better} to calculate ROUGE-WE. Both $\textrm{clef}\_\textrm{inex}_{1gram}$ and $\textrm{clef}\_\textrm{inex}_{2gram}$ models where trained with the $34,507$ passages of the TC@INEX 2012 task. Finally, the $\textrm{wiki}_{2gram}$ model was trained with a $3.4Gb$ partition of the 2012's Wikipedia.

\subsection{Meta Evaluation}

Rather than calculating the correlation between systems' average measure scores and their human assigned average coverage scores as in \cite{DBLP:conf/naacl/LinH03} with a limited number of systems and abstracts, we considered each passage individually for which we have a human assessment. We then calculated all scores for each metric.

Following the same approach as in TREC 2015 Microblog Track\footnote{\url{https://github.com/lintool/twitter-tools/wiki/TREC-2015-Track-Guidelines}} \cite{yilmaz2015overview}, to evaluate one specific metric we ranked passages in decreasing order following this metric. Considering different cut-off values, we computed the normalized Cumulative Gain (nCG) over top ranked passages. This score is the sum of the graded human judgments top ranked passages obtained in the TC@INEX 2012 divided by the maximum score that could have been expected at this precise cut-off value.

More precisely, given a set of topics $\cal T$ and a set of passages $\Omega$ for which there is a graded evaluation $ref_\tau(\omega)$ of their Informativeness for at least one topic $\tau \in {\cal T}$, $nCG_k(m)$ is computed as follows for any metric $m$ and any cut-off value $k$:
\begin{equation}
nCG_k(m)=\frac{\max\{\sum_{\omega\in S} m(\omega):S\subset \Omega, |S|\leq k\}}
{\max\{\sum_{\omega\in S} ref_\tau(\omega):S\subset \Omega, |S|\leq k, \tau \in {\cal T}\}}
\end{equation}

\noindent where $|S|$ is the cardinal of $S$.

For values of $k$ lower than the number of passages $\omega$ such that $ref_\tau(\omega)\geq 1$, $nCG_k(m)$ shows the measure Precision. On the contrary, for cut-off values higher than the number of passages $\omega$ such that $ref_\tau(\omega)>0$,  $nCG_k(m)$ indicates the maximum Recall that can be expected using this measure.

We considered and evaluated the following set ${\cal M}$ of 13 textual overlap measures:
\begin{description}
\item[F1\_1:] F1-score among uni-grams
\item[F1\_2:] F1-score among bi-grams
\item[F1\_sk:] F1-score among skip-grams with a gap of one word
\item[KL\_1:] KL divergence among uni-grams
\item[KL\_2:] KL divergence among  bi-grams
\item[KL\_sk:] KL divergence among skip-grams with a gap of one word
\item[LS\_1:] LogSim score among uni-grams
\item[LS\_2:] LogSim score among bi-grams
\item[LS\_sk:] LogSim score among skip-grams with a gap of one word
\item[w2v\_g:] Word2vec cosine similarity over the Google News model ($\textrm{google}\_\textrm{news}_{1gram}$)
\item[w2v\_c:] Word2vec cosine similarity over the set of all passages in $\Omega$ ($\textrm{clef}\_\textrm{inex}_{1gram}$)
\item[w2v\_c\_bi:] Word2vec cosine similarity the same set $\Omega$ but considering word bi-grams instead of single words ($\textrm{clef}\_\textrm{inex}_{2gram}$)
\item[w2v\_wp\_bi:] Word2vec cosine similarity over the Wikipedia version used as copora in TC@INEX 2012 considering word bi-grams ($\textrm{wiki}_{2gram}$)
\end{description}

The first nine measures are discrete measures that we also applied to sibling operators over Wikipedia anchor texts.
We consider the anchors associated with Wikipedia entities as potential nuggets as defined in Pyramid evaluations \cite{dang2005overview}.

To evaluate Informativeness, each measure was applied to estimate the overlap between the reference informative passages of the topic and the passage itself.
Statistical signification was tested over topics.
Meanwhile, to evaluate Interestingness each measure is applied to estimate the overlap between the set of all informative passage, disregarding its specific topic and the passage itself. Statistical signification is tested based on a 12-fold split of the corpus. 

\section{Results}
\label{sec:Results}
\subsection{Which Measure Ranks First Most Informative Passages per Topic?}

We used nCG over TC@INEX 2012 data to test the ability of state of the art Informativeness measures to evaluate ATS systems.
The goal is to distinguish informative from non-informative short text passages rather than entire abstracts.
This is done by ranking the Informativeness score of all passages manually assessed in TC@INEX 2012 for any topic and sorting them in decreasing order. Then, the ability of a specific measure to find the $k$ most informative passages is evaluated.

For each measure $\mu\in{\cal M}$ and  passage $\omega\in\Omega$ associated with a topic $\tau \in {\cal T}$, we compute  $\mu(\varphi_\tau,\omega)$ to estimate the semantic overlap between $\omega$ with the textual reference $\varphi_\tau$ defined as: 
\begin{eqnarray}
\varphi_\tau&=&\bigcup\{\omega\in\Omega: ref_\tau(\omega)>0\}
\end{eqnarray}

\subsection*{Discrete vs. Continuous Metrics}

\begin{figure}[h]
\begin{center}
\includegraphics[trim={2cm 1.5cm 2cm 1cm},clip=true,width=\columnwidth]{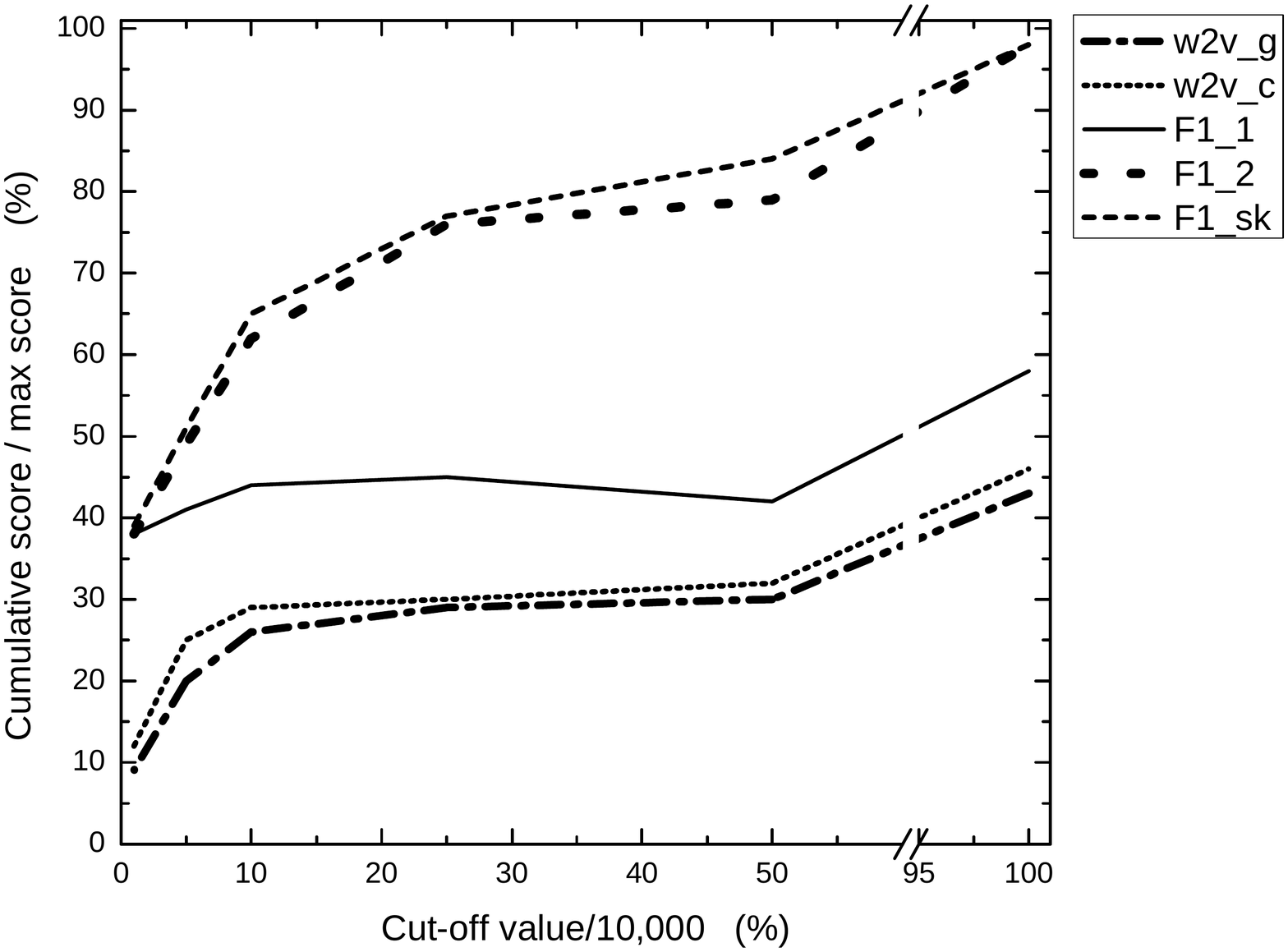}
\caption{\label{fig:infF1we}Informativeness nCG scores about F1-scores vs. word2vec approaches}
\end{center}
\end{figure}

Figure \ref{fig:infF1we} presents the results of the F1-scores over uni-grams, bi-grams and skip-grams and compares them against the word2vec uni-gram models.
It shows that word2vec approaches are less efficient to evaluate short passage informativeness even over standard uni-grams.
Meanwhile, F1-scores over bi-grams and skip-grams reach similar performance.

The figure also shows that for low cut-off values, Precision is lower that $40$\% and that bi-grams do not improve the measure performance.
For higher cut-off values it shows that maximal Recall is only reached by the F1-score over bi-grams and skip-grams.

A slightly improvement can be seen with the local word2vec model ($\textrm{clef}\_\textrm{inex}_{1gram}$) over the model trained with the Google News dataset ($\textrm{google}\_\textrm{news}_{1gram}$).
This behavior can be explained by the fact that the amount of unknown words in the $\textrm{clef}\_\textrm{inex}_{1gram}$ model is smaller that the one in the $\textrm{google}\_\textrm{news}_{1gram}$ model.
Despite that this improvement in the cumulative score is very small, it shows that is better to create the word embeddings with a smaller but more specialized dataset.

\subsection*{Best Discrete informativeness Metric}

\begin{figure}[h]
\begin{center}
\includegraphics[trim={2cm 1.1cm 2cm 1cm},clip=true,width=\columnwidth]{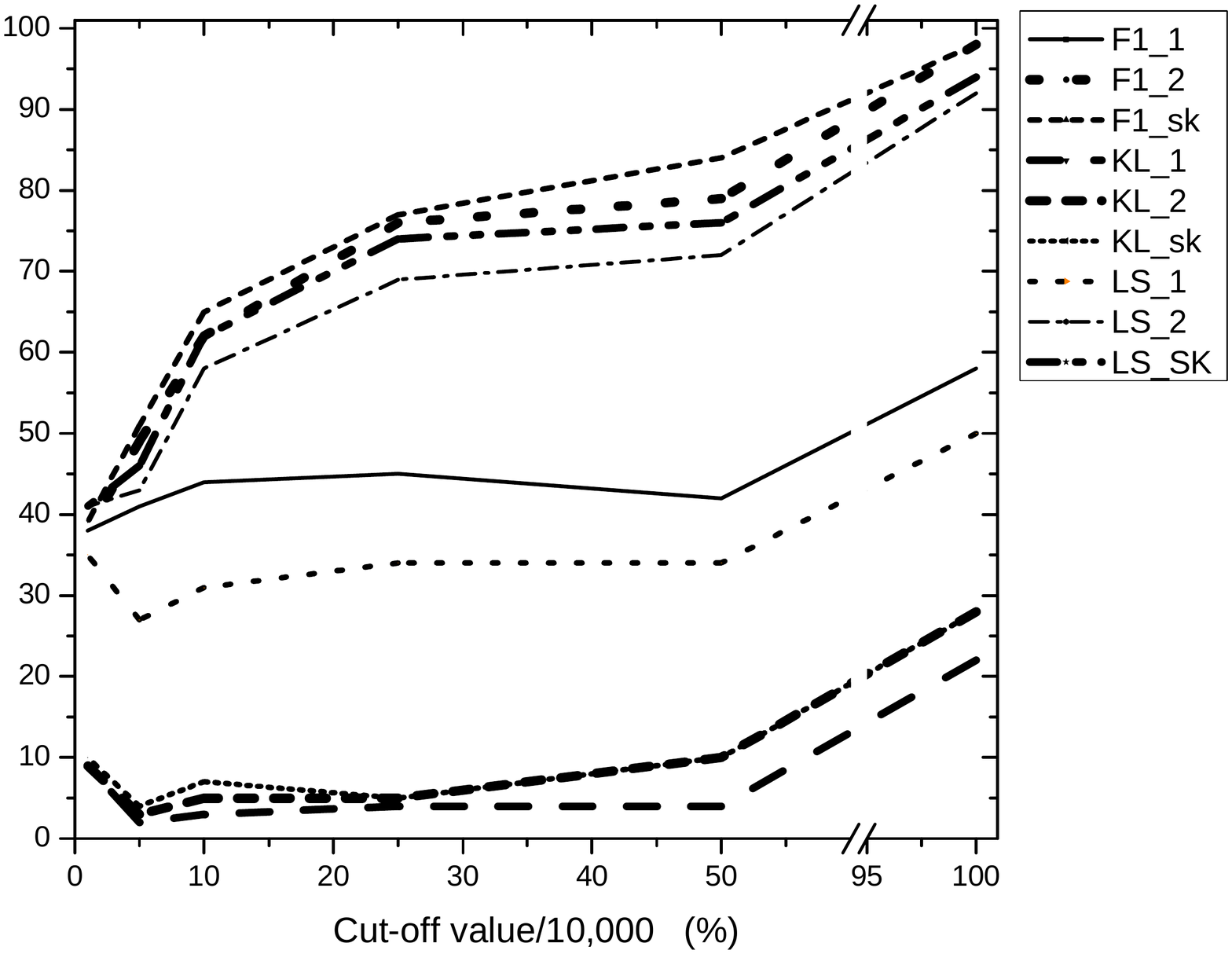}
\caption{\label{fig:infLS}Informativeness nCG scores about F1-scores vs. KL and LogSim approaches}
\end{center}
\end{figure}

In Figure \ref{fig:infLS} it can be seen that all KL metrics show similar performance for medium cut-off values.
Among all metrics over uni-grams, F1-score achieves the better Precision and Recall over all cut-off values, obtaining a pick in the recall of almost $60$\% in high cut-off values.
F1-scores over bi-grams and skip-grams outperforms LogSim over the same units, however this improvement is not significant.
Again only bi-gram based measures reach maximal recall for high cut-off values. 

\subsection*{Discrete Metrics Restricted to Entities}

\begin{figure}[h!]
\begin{center}
\includegraphics[trim={1cm 1.5cm 0.5cm 0.5cm},clip=true,width=\columnwidth]{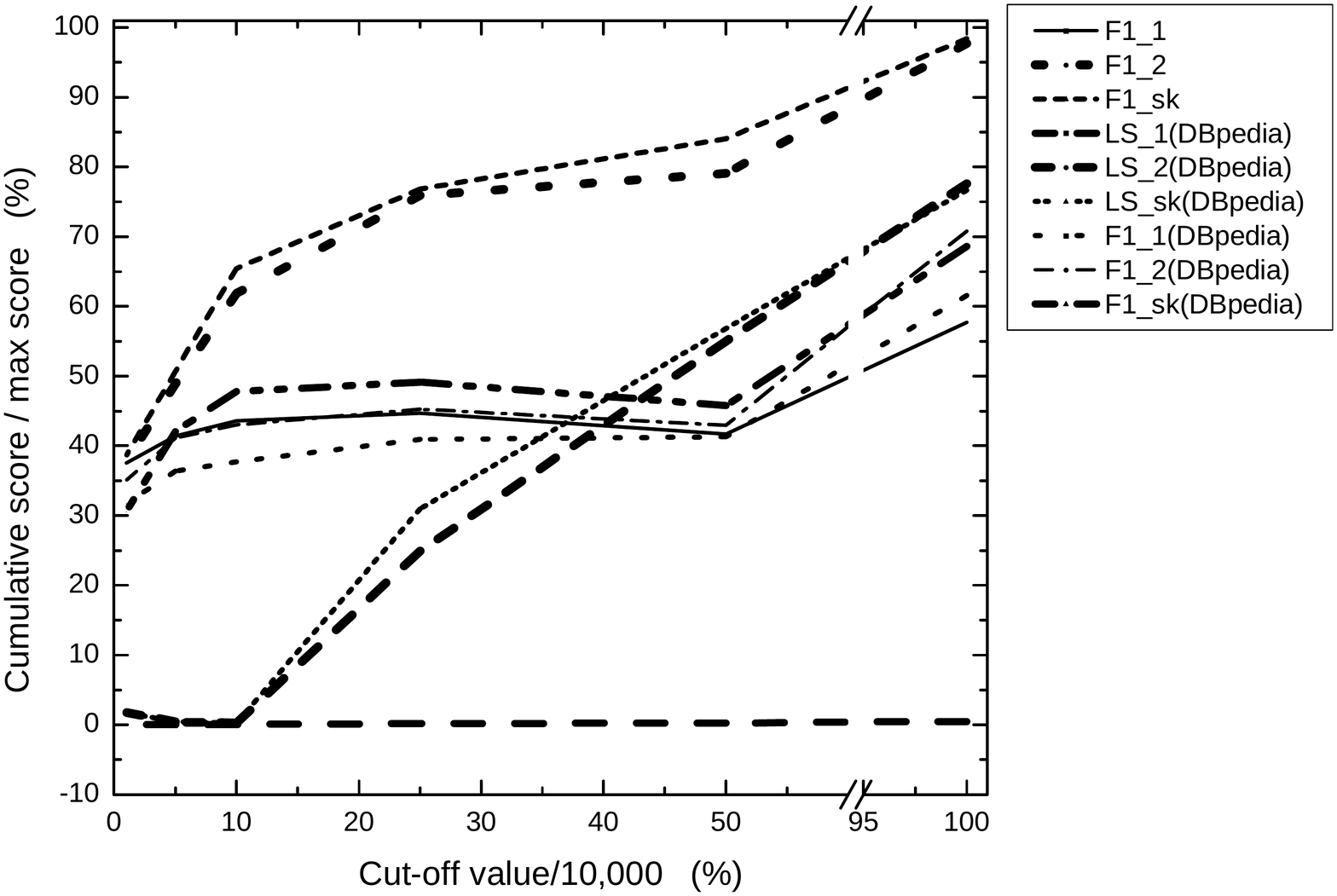}
\caption{\label{fig:infent}Informativeness nCG scores about DBpedia entities vs. complete passages}
\end{center}
\end{figure}

As shown in Figure \ref{fig:infent}, it appears that for low cut-off values, restricting references and passages to DBpedia entities provides equivalent scores to complete passages for F1-scores.
By contrast, for high cut-off values and maximal recall there is a clear gap between restricted entities and complete passages.

An interesting remark is the behavior of the LogSim measures with DBpedia entities.
For low cut-off values all three LogSim scores are extremely low and fail to highlight informative passages; for high cut-off values, LogSim scores over bi-grams and skip-grams reach a recall near the $70\%$.

In regard to the F1-scores with uni-grams, a slightly outperform with restricted entities over complete passages can be seen for high cut-off values but this difference is not significant.

\subsection{Which Measure Ranks First Most Interesting Passages?}

We now used nCG over TC@INEX 2012 data to test the ability of the same Informativeness measures in distinguishing interesting from non-interesting text passages without considering an explicit topic.

For each measure $\mu\in{\cal M}$ and passage $\omega\in\Omega$ associated with a topic $\tau \in {\cal T}$, we compute  $\mu(\Delta_\tau,\omega)$ to estimate the semantic overlap between $\omega$ and the textual reference $\Delta_\tau$ defined as the concatenation of passages that are informative for at least one different topic $t^\prime\neq t$: 

\begin{eqnarray}
\Delta_\tau&=&\bigcup\{\omega\in\Omega: (\exists t^\prime\in{\cal T}-\{t\}) ref_{t^\prime}(\omega)>0\}
\end{eqnarray}

However in oder to avoid any overfitting effect, we've split the dataset of $34,507$ passages ranked per topic into $12$ folds and restricted $\Delta_\tau$ to passages in a different fold than the one including $\tau$.

\subsection*{Discrete vs. Continuous Metrics}

\begin{figure}[h!]
\begin{center}
\includegraphics[trim={2cm 1.5cm 0cm 1cm},clip=true,width=\columnwidth]{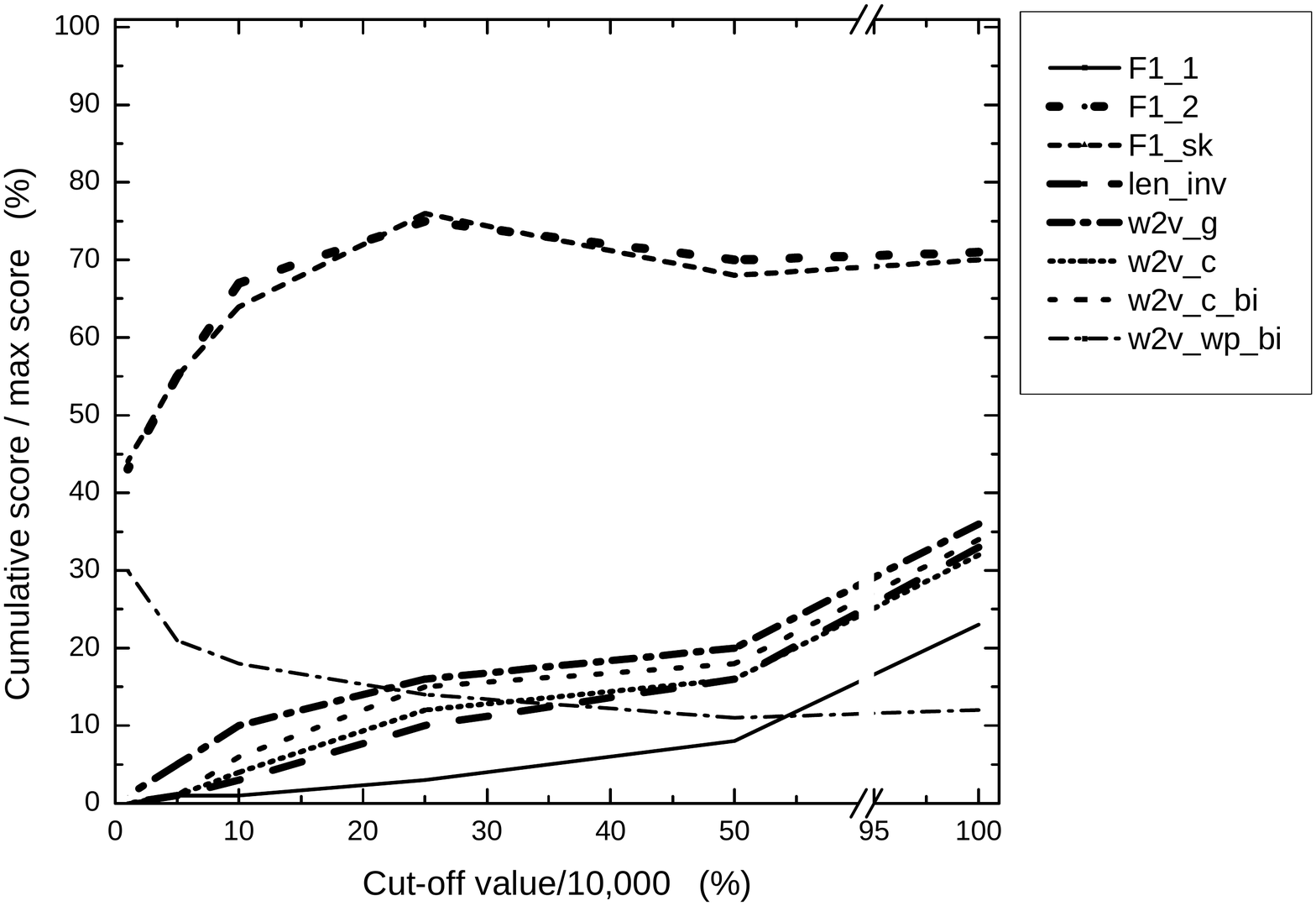}
\caption{\label{fig:intwe}Interestingness nCG scores about F1-scores vs. word2vec approaches}
\end{center}
\end{figure}

\noindent Figure \ref{fig:intwe} presents the results of F1 scores and word2vec models over uni-grams and bi-grams.
For small cut-off values, word2vec models and the F1-score over uni-grams show a similar low performance.
The improvement of F1-scores over bi-grams and skip-grams against uni-grams for all cut-off values was striking for this task. 

F1-scores do not reach this time maximal recall for high cut-off values.
However best compromise between recall and precision is reached for a cut-off value around $2,500$ passages over $3,000$ informative passages in the reference.
Word2vec models trained over Wikipedia using bi-grams significantly improves precision over small cut-off values but then become less efficient after $2,500$ passages because of the proportion of missing bi-grams in the model.
Other word2vec models do not outperform the baseline taking inverse passage length (len\_inv) as measure with the idea that short passages could be more specific. 

\subsection*{Best Discrete Interestingness Metric}

\begin{figure}[h]
\begin{center}
\includegraphics[trim={1.5cm 1.5cm 2cm 1cm},clip=true,width=\columnwidth]{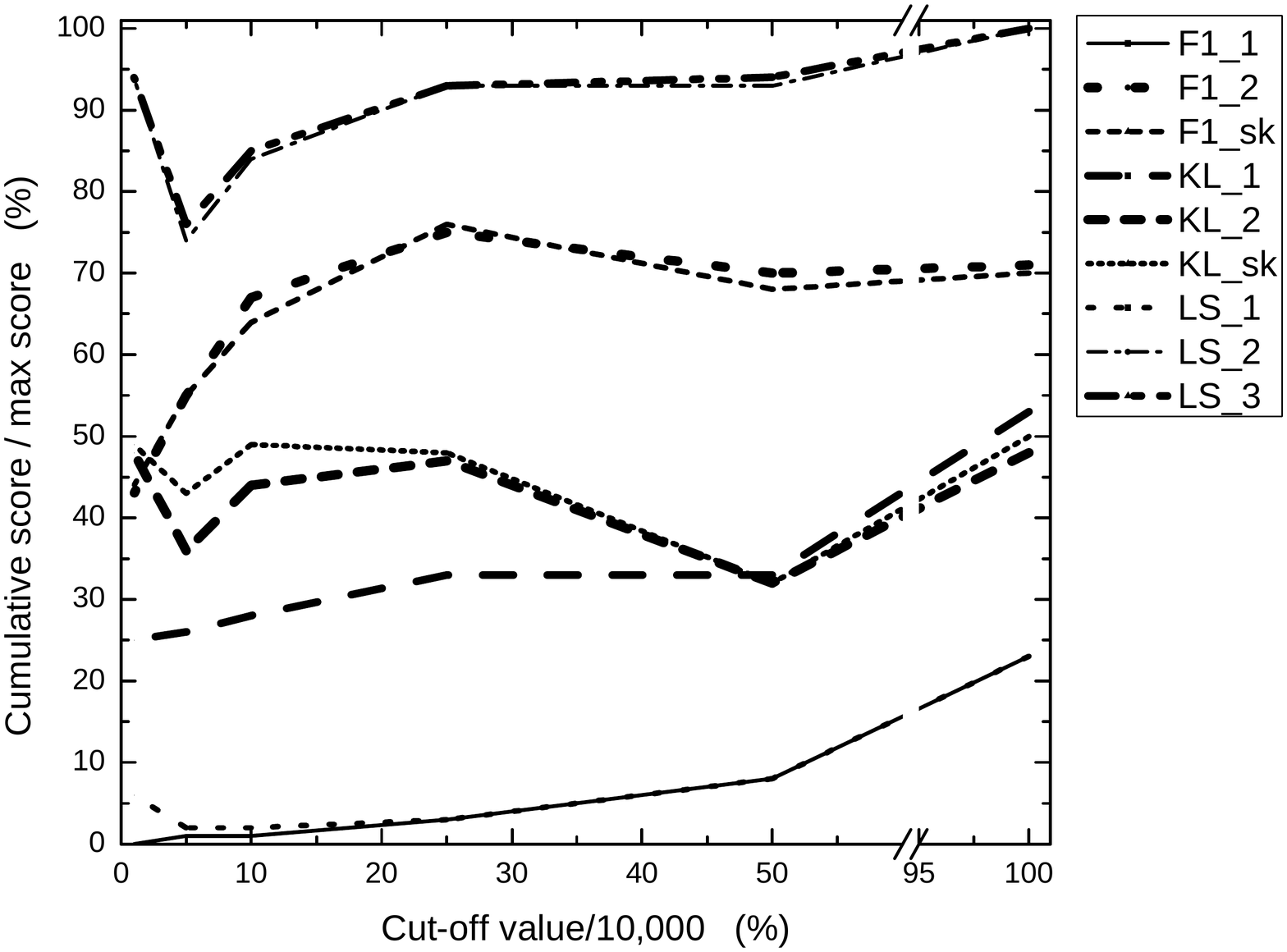}
\caption{\label{fig:intLS}Interestingness nCG scores about F1-scores vs. KL and LogSim approaches}
\end{center}
\end{figure}

As shown in Figure \ref{fig:intLS}, among discrete measures there is another contrast with previous results over Informativeness.
It appears that LogSim measures over bi-grams significantly outperform F1-scores and reach total recall after $5,000$ retrieved passages.
Again, only bi-gram and skip-gram based measures reach maximal recall for high cut-off values. LogSim bi-gram and skip-gram measures significantly outperform all other measures for any cut-off value.
It also appears that KL metrics provide the best results over uni-grams among all metrics with the same units.

\subsection*{Discrete Metrics Restricted to Entities}

\begin{figure}[h!]
\begin{center}
\includegraphics[trim={2cm 1.5cm 0cm 0cm},clip=true,width=\columnwidth]{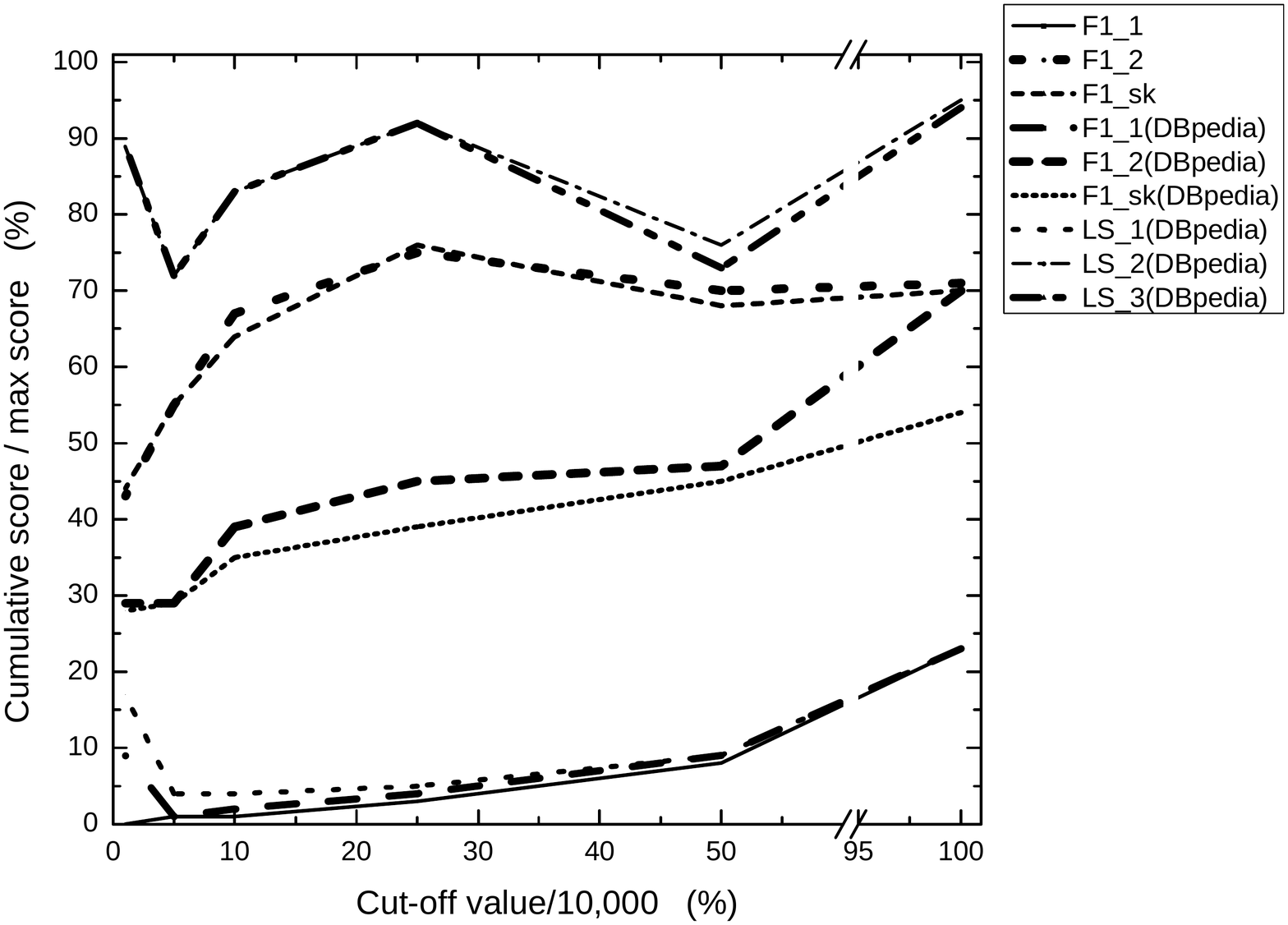}
\caption{\label{fig:intent}Interestingness nCG scores about DBpedia entities vs. complete passages}
\end{center}
\end{figure}

Finally, in Figure \ref{fig:intent} we look at the impact of restricting references and passages to DBpedia entities in the case of interestingness.
For low and medium cut-off values, F1-scores over bi-grams and skip-grams with complete passages show a better performance than the ones with DBpedia entities. This difference in performance is reduced for high cut-off values where the recall for bi-grams with DBpedia entities rises to a $70\%$. 

From Figures \ref{fig:intLS} and \ref{fig:intent} we see that LogSim measures over bi-grams and skip-grams with both complete  and restricted passages outperform all the other measures.
In general, LogSim scores with complete passages show a better performance than those with restricted entities. Surprisingly, it appears that LogSim measures over bi-grams remain almost stable and outperform F1-scores both over complete and restricted passages. 

\section{Conclusion}
\label{sec:Conclusion}

In this paper we have defined the concept of Interestingness in FR as a generalization of the concept of Informativeness where the information need is diverse and formalized as an unknown set of implicit queries.

We then studied the ability of state of the art Informativeness measures to cope with this generalization, showing that on this new framework, cosine similarity between word embedding vector representations outperform discrete measures only on uni-grams.
Moreover discrete bi-gram and skip-gram LMs appeared to be a key point of Interestingness evaluation, significantly outperforming all experiments wit word2vec models. 
However we did show that an alternative word2vec bi-gram model learned on Wikipedia outperforms the uni-gram word2vec models on nCG scores for small cut-off value, but its performance decreases for higher values.
Finally we showed that TC@INEX 2012 LogSim measure provides indeed best results for efficient Interestingness detection over this corpus, both on complete passages and on passages restricted to their anchor texts referring to entities.

\bibliographystyle{splncs03}
\bibliography{mybibfile.bib}

\begin{thebibliography}{10}
\providecommand{\url}[1]{\texttt{#1}}
\providecommand{\urlprefix}{URL }

\bibitem{bellot2013overview}
Bellot, P., Moriceau, V., Mothe, J., Sanjuan, E., Tannier, X.: Overview of the
  inex 2013 tweet contextualization track. In: CLEF 2013 Evaluation Labs and
  Workshop, Online Working Notes (2013)

\bibitem{DBLP:journals/ipm/BellotMMST16}
Bellot, P., Moriceau, V., Mothe, J., SanJuan, E., Tannier, X.: {INEX Tweet
  Contextualization task: Evaluation, results and lesson learned}. Information
  Processing and Management  52(5),  801--819 (2016)

\bibitem{dang2005overview}
Dang, H.T.: {Overview of DUC 2005}. In: {Proceedings of the Document
  Understanding Conference}. pp. 1--12 (2005)

\bibitem{dang2007overview}
Dang, H.T., Kelly, D., Lin, J.J.: Overview of the trec 2007 question answering
  track. In: TREC. vol.~7, p.~63 (2007)

\bibitem{DBLP:conf/sigir/Ekstrand-AbuegPA13}
Ekstrand-Abueg, M., Pavlu, V., Aslam, J.A.: Live nuggets extractor: a
  semi-automated system for text extraction and test collection creation. In:
  Jones, G.J.F., Sheridan, P., Kelly, D., de~Rijke, M., Sakai, T. (eds.) SIGIR.
  pp. 1087--1088. ACM (2013)

\bibitem{DBLP:conf/inex/2011}
Geva, S., Kamps, J., Schenkel, R. (eds.): Focused Retrieval of Content and
  Structure, 10th International Workshop of the Initiative for the Evaluation
  of {XML} Retrieval, {INEX} 2011, Saarbr{\"{u}}cken, Germany, December 12-14,
  2011, Revised Selected Papers, Lecture Notes in Computer Science, vol. 7424.
  Springer (2012)

\bibitem{harris1954distributional}
Harris, Z.S.: Distributional structure. Word  10(2-3),  146--162 (1954)

\bibitem{Hilderman:2003}
Hilderman, R.J., Hamilton, H.J.: Measuring the interestingness of discovered
  knowledge: A principled approach. Intell. Data Anal.  7(4),  347--382 (2003)

\bibitem{DBLP:journals/sigir/KampsGT08}
Kamps, J., Geva, S., Trotman, A.: Report on the {SIGIR} 2008 workshop on
  focused retrieval. {ACM SIGIR} Forum  42(2),  59--65 (2008)

\bibitem{koh_interestingness_2008}
Koh, Y.S., O'Keefe, R., Rountree, N.: Interestingness {Measures} for
  {Association} {Rules}: {What} {Do} {They} {Really} {Measure}? In: Data Mining
  and Knowledge Discovery Technologies, chap.~2, pp. 36--58. IGI PUBLISHING
  (2008)

\bibitem{Lin2004Rouge}
Lin, C.Y.: Rouge: a package for automatic evaluation of summaries. pp. 74--81
  (July 25-26 2004)

\bibitem{DBLP:conf/naacl/LinH03}
Lin, C.Y., Hovy, E.H.: {Automatic Evaluation of Summaries Using N-gram
  Co-occurrence Statistics}. In: HLT-NAACL. vol.~1, pp. 71--78 (2003)

\bibitem{lin2004looking}
Lin, C.Y., Och, F.: Looking for a few good metrics: Rouge and its evaluation.
  In: NTCIR Workshop (2004)

\bibitem{DBLP:conf/sigir/LinZ07}
Lin, J.J., Zhang, P.: Deconstructing nuggets: the stability and reliability of
  complex question answering evaluation. In: Kraaij, W., de~Vries, A.P.,
  Clarke, C.L.A., Fuhr, N., Kando, N. (eds.) SIGIR. pp. 327--334. ACM (2007)

\bibitem{LouisN:09}
Louis, A., Nenkova, A.: {Performance Confidence Estimation for Automatic
  Summarization}. In: EACL. pp. 541--548. ACL (2009)

\bibitem{Mikolov2013_Efficient}
Mikolov, T., Chen, K., Corrado, G., Dean, J.: Efficient estimation of word
  representations in vector space. CoRR  abs/1301.3781 (2013)

\bibitem{mikolov2013distributed}
Mikolov, T., Sutskever, I., Chen, K., Corrado, G.S., Dean, J.: Distributed
  representations of words and phrases and their compositionality. In: Advances
  in neural information processing systems. pp. 3111--3119 (2013)

\bibitem{mikolov2013linguistic}
Mikolov, T., Yih, W.t., Zweig, G.: Linguistic regularities in continuous space
  word representations. In: HLT-NAACL. vol.~13, pp. 746--751 (2013)

\bibitem{morin2005hierarchical}
Morin, F., Bengio, Y.: Hierarchical probabilistic neural network language
  model. In: Proceedings of {Aistats}. vol.~5, pp. 246--252 (2005)

\bibitem{ng2015better}
Ng, J.P., Abrecht, V.: Better summarization evaluation with word embeddings for
  rouge. arXiv:1508.06034 [cs.CL]  (2015)

\bibitem{DBLP:conf/wsdm/PavluRGA12}
Pavlu, V., Rajput, S., Golbus, P.B., Aslam, J.A.: Ir system evaluation using
  nugget-based test collections. In: Adar, E., Teevan, J., Agichtein, E.,
  Maarek, Y. (eds.) WSDM. pp. 393--402. ACM (2012)

\bibitem{DBLP:conf/acl/RadevTSLBQCLD03}
Radev, D.R., Teufel, S., Saggion, H., Lam, W., Blitzer, J., Qi, H., \c{C}elebi,
  A., Liu, D., Dr{\'a}bek, E.: {Evaluation Challenges in Large-Scale Document
  Summarization}. In: ACL'03. pp. 375--382 (2003)

\bibitem{rong2014Word2vec}
Rong, X.: {word2vec Parameter Learning Explained}. arXiv:1411.2738 [cs.CL]
  (2014)

\bibitem{DBLP:conf/coling/SaggionMCSV10}
Saggion, H., Torres-Moreno, J.M., da~Cunha, I., SanJuan, E.,
  Vel{\'a}zquez-Morales, P.: {Multilingual Summarization Evaluation without
  Human Models}. In: Huang, C.R., Jurafsky, D. (eds.) COLING. pp. 1059--1067
  (2010)

\bibitem{DBLP:conf/inex/SanJuanMTBM11}
SanJuan, E., Moriceau, V., Tannier, X., Bellot, P., Mothe, J.: Overview of the
  {INEX} 2011 question answering track (qa@inex). In: Geva et~al.
  \cite{DBLP:conf/inex/2011}, pp. 188--206

\bibitem{DBLP:conf/clef/SanJuanMTBM12}
SanJuan, E., Moriceau, V., Tannier, X., Bellot, P., Mothe, J.: {Overview of the
  INEX 2012 Tweet Contextualization Track}. In: Forner, P., Karlgren, J.,
  Womser-Hacker, C. (eds.) CLEF (Working Notes/Labs/Workshop) (2012)

\bibitem{Torres2014}
Torres-Moreno, J.M.: Automatic Text Summarization. John Wiley and Sons, London
  (2014)

\bibitem{yilmaz2015overview}
Yilmaz, E., Verma, M., Mehrotra, R., Kanoulas, E., Carterette, B., Craswell,
  N.: Overview of the trec 2015 tasks track. In: {Text Retrieval Conference
  (TREC)} (2015)

\end{thebibliography}

\end{document}